\documentclass{amsart}
\usepackage{amssymb}
\def\preprint{}       
\def\finished{}
\def\archive {}
\newcommand{\cqd}{\hfill $\rule{2mm}{2mm}$}

\newtheorem{define}{Definition}        
\newtheorem{lemma}{Lemma} 
\long\def\abstract{ This works deals with the formal mathematical structure of so called Grassmann Numbers applied to Theoretical Physics, which is a basic concept on Berezin integration. To achieve this purpose we make use of some constructions from relative modern Polynomial Identity Algebras (PI-Algebras) applied to the special case of the Grassmann algebra. \\
\begin{description}
	\item[{\bf Key-words}] Abstract Algebra, Grassmann Numbers, Theoretical Physics, Non-commutative Algebras.	
\end{description}
}
\input{epsf}
\begin{document}
\vskip 10mm
{\hfill \archive   \vskip -2pt \hfill\preprint }
\vskip 10mm
\centerline{\huge\bf Building Grassmann Numbers}
\vskip 4mm
\centerline{\huge\bf from PI-Algebras.}
\vskip 7mm
\begin{center} \vskip 7mm
Ricardo Bent\'{\i}n{\begingroup\def\thefootnote{$\aleph$}
                                \footnote{e-mail: rbentin@uesc.com}
                                \addtocounter{footnote}{-1} \endgroup}
and
Sergio Mota{\begingroup\def\thefootnote{$\beth$}
                                \footnote{e-mail: smalves@uesc.br}
                                \addtocounter{footnote}{-1} \endgroup}\\                           
 Departamento  Ciências Exatas e Tecnológicas - DCET\\
 Universidade Estadual de Santa Cruz\\
 Rodovía Ilh{\'e}us-Itabuna km 16 s/n, CEP 45662-000, Ilh{\'e}us-BA, Brazil.
 {\begingroup\def\thefootnote{}
                                \footnote{\tiny{erev Pesach 5772}}
                                \addtocounter{footnote}{-1} \endgroup}    
\vfill {\bf ABSTRACT} \vskip 3mm \end{center}    \abstract
\vfill \noindent \\[5pt] \finished \vspace*{3mm}
\thispagestyle{empty} \newpage
\pagestyle{plain}
\newpage
\setcounter{page}{1}
\section{\sc Introduction.}
The ideas of H. Grassmann in Mathematics and later F. Berezin in Physics\cite{pioneers} are a good example of how science fields required Mathematics to obtain new results. But as any other field, Mathematics evolves. Follow on this philosophy, we make use of some kind of new Mathematics to examine the {\it Grassmann Numbers}, now using the concepts of Polynomial Identity Theory (PI-Theory), that is part of non-commutative algebras.\\
The misunderstanding arises in the following polynomial on Berezin's work:
\begin{equation}
\label{nn}
p(\theta)=a+b\theta,
\end{equation}
that according to usual polynomial definition over some field $K$, the coefficients $a,b$ must belong to the same field, this means that $a,b\in K$. But Berezin's work states that $a\in\mathbb{C}$ (complex number) and $b$ is a kind of C-number, since anti-commutation is required for $b$ and $\theta$. This apparent paradox is the central motivation of this work, which is a result of discussions of the authors over some earlier works\cite{us}. We divided this work as: Next section we make a brief review of PI-Algebras concepts. After a section on Mathematical background, we start a new section on Physics background, where we will describe how the Berezin's ideas appears on the context of Supersymmetry\cite{super}. The next section we will joins the ideas discussed in previous sections focusing on the above polynomial.\\
The final section contains our conclusions.
\section{\sc PI-Algebras Fundamentals.}
In this section we will present some introductory ideas and the basic Mathematics we required to tackle this problem. This sections is written in order Theoretical Physicists can understand the main ideas of PI-Algebras. We will started with some historical motivation and with basics definitions on non-commutative algebra, after that we will give some examples of polynomial identities, and finally give, as an example, the case of Grassmann algebras from this point of view.\\
In 1987 Kemer \cite{kemer} showed that T-prime algebras (i.e., algebras with T-ideals that are T-prime) that are non-trivial (in characteristic zero) are of type $M_n(K)$, $M_n(\mathbb{G})$ and $M_{a,b}(\mathbb{G})$, where $\mathbb{G}$ is the finite dimensional Grassmann algebra over a field $K$.\\
After a short introductory, now we are going to give some definitions \cite{drensky}. The main structure to study is an algebra, then it is defined as
\begin{define}[Algebra]
	\label{algebra} A vector space $A$ ($a,b,c\in A$) over a field $K$ ($\alpha\in K$) with a binary operation on $A$ defines an algebra if and only if:
	\begin{itemize}
	\item $(a+b)c=ac+bc$,
	\item $a(b+c)=ab+bc$, and
	\item $\alpha(ab)=(\alpha a)b=a(\alpha b)$.
	\end{itemize}
\end{define}
Some times this kind of algebra is also called \textit{K-Algebra}\\
Now lets introduce some non-commutativity. Let $$X=\{x_1,x_2,\cdots,x_n\}$$ be a set, where each $x_i$ must obey in general the property $x_ix_j\neq x_jx_i,\ \forall i,j:i\neq j$, and of course this last condition implies there exist some binary operation among the $x_i$'s.\\ 
Let's introduce another binary operation, and denote this by $(+)$, altogether with the previous one, allows us to built up elements of type, for example, 
$$x_ix_jx_k+x_jx_kx_i+x_kx_ix_j,$$ $$x_ix_j+x_jx_k,$$ etc., that according the mandatory definition of binary operation, must also belongs to the main set $X$. 
Using the field $K$ and the elements of $X$ we can make the following construction,
$$B=\{1,x_{i_1}\cdots x_{i_k}:\ i_l=1,2\cdots\},$$ then we have a vector space $V$ generated by $B$ over $K$, and also $K\langle X\rangle$ is the associative algebra generated by the vector space $V$ where the the binary operation is now defined as:
$$(x_{i_1}\cdots x{i_k})(x_{j_1}\cdots x{j_l})=x_{i_1}\cdots x{i_k}x_{j_1}\cdots x{j_l}.$$ 
We will call the elements of $K\langle x\rangle$ as polynomials.
\begin{define}[right-ideal]
	\label{ideal} Let $A$ be an algebra, and $I$ a sub-algebra of $A$, then $I$ represents an right-ideal if and only if for all $a\in A$ and $i\in I$ we have $ai\in I$
\end{define}
In the same form we define a left-ideal.
\begin{define}[2-ideal] An ideal $I$ of an algebra $A$ is said a two-ideal if it is both, right-ideal and also left-ideal.
	\label{2-ideal} 
\end{define}
\begin{define}[homomorphism] A linear transformation $\varphi:A\rightarrow B$ between the algebras $A$ and $B$ is said to be an homomorphism is and only if $$\varphi(ab)=\varphi(a)\varphi(b),$$
in few words, operations are preserved.
	\label{homomorphism} 
\end{define}
If the above definition is applied over the same set, the homomorphism is called {\it endomorphism}.
\begin{define}[T-ideal] An ideal $I$ of an algebra $A$ is defined as a T-ideal if $I$ is invariant under all endomorphisms on $A$, i.e., $\Phi(I)\subseteq I$ for all $\Phi:A\rightarrow A$. 
	\label{T-ideal} 
\end{define}
Our construction requires another important definition:
\begin{define}[Quotient Algebra]
	\label{quotient} Let $I$ be an 2-Ideal of 
	$K\langle x\rangle$ and $f,g\in K\langle x\rangle$, 
	then we define the quotient algebra over $I$ as $K\langle x\rangle/I$ where as usual
	\begin{itemize}
	\item $(f+I)+(g+I)=(f+g)+I$,
	\item $(f+I)(g+I)=fg+I$.
	\end{itemize}
	Notice that $\bar{f}=f+I\in K\langle x\rangle/I$
\end{define}
Now we are ready to define the Grassmann algebra $\mathbb{G}$ \cite{plamen}:
\begin{define}[Grassmann Algebra] Let be a field $K$, a non-commutative algebra $A$ over $K$, and the ideal $I=I=\langle x_ix_j+x_jx_i\rangle,\ i,j\geq 1$, the quotient algebra $$\mathbb{G}=A/I.$$
	\label{Grassmann Algebra} 
\end{define}
According to this definition it is not hard to see the Grassmann algebra can be written as:
$$\mathbb{G}=alg_K\{1,x_{i_1}x_{i_2}\cdots x_{i_n};\ i_1<i_2<\cdots<i_n\}.$$
Another important definition is concerned with the {\it commutative} sub-algebra of an algebra $A$ that is called the center of the algebra and is defined as:
\begin{define}[Center of an Algebra] 
Let $A$ be a non-commutative algebra, then the sub-set $Z[A]=\{x\in A:\forall g\in A,\ gxg^{-1}=x\}$ is defined as the center of the algebra $A$.
	\label{center}
\end{define}
Using this last definition, we can obtain two sub-spaces of $\mathbb{G}$:
$$\mathbb{G}_0=alg_K\{1,x_{i_1}x_{i_2}\cdots x_{i_n};\ i_1<i_2<\cdots<i_n,\ n\ even\}$$
$$\mathbb{G}_1=alg_K\{x_{i_1}x_{i_2}\cdots x_{i_n};\ i_1<i_2<\cdots<i_n,\ n\  odd\}$$
Where $\mathbb{G}_0$ is the center of $\mathbb{G}$, this means $\mathbb{G}=\mathbb{G}_0\oplus\mathbb{G}_1$.\\
Let us to give one more definition:
\begin{define}[Graded Algebra] 
Let $A$ have the structure of a non-commutative algebra, if $A=\oplus_{g\in G}A_g$, where $A_g$ is a subspace of $A$ ($\forall g\in G$), and ($\forall g,h\in G$), $A_gA_h\subseteq A_{g+h}$, then $A$ is called a G-graded algebra.
	\label{graded}
\end{define}
According this definition, the Grassmann algebra is also a 2-Graded algebra.
Here we stop our succinct description of PI-Theory.
\section{\sc Physics Behind these Mathematical Concepts.}
Trying to understand the fundamental structure of Nature is not an easy task. For many years researchers have been working hard to obtain a better understanding of this problem, and Mathematics, a trust discipline, was (and is) a mandatory machine giving assistance on this kind of work for many years, this is because, in a few words, Physics is a science that makes use of Mathematics trying to describe Nature.\\
Fist we will describe in few lines, the main physical ideas, in order mathematicians figure out the Physics behind these concepts. It is not our aim to give a full review on this topic, for that purpose there are good works on it \cite{physics}.
One interesting fact in Nature is that Nature itself is split in two main disjoint sets (mathematically speaking, we are saying two equivalence classes):\\
Fermions, obeying the anti-commuting algebra: 
$$\{b,b^\dag\}=1,$$ 
and Bosons, that observe the commuting algebra:
$$[a,a^\dag]=1.$$ 
This historical fact (associated with statistical distributions) also gives support to understand main concepts in Elementary Particle Physics. In our opinion, a good candidate for describing the fundamental interactions of nature is superstring theory, that uses the ideas of super-symmetry\cite{super}, where the Grassmann algebra and Berezin integration\cite{pioneers} are important and make use polynomials of type:
$$p(\theta)=a+b\theta.$$
As states above, according the basics text of algebra, the coefficients $a$ and $b$ must belongs to the same field, say it $K$, but according to Theoretical Physics, $a$ belongs to the complex field and $b$ to the so called Grassmann Numbers. On the following section we will work on this apparent paradox.\\
Summing up, the particle physics classifications as understand it, today from a (quantum) statistical distribution point of view is: we have some particles called Bosons (in memory of S. E. Bose and A. Einstein), and some other particles called Fermions (in memory of E. Fermi and P. A. M. Dirac).\\
It is at this point that Grassmann algebra arises.
\section{\sc Grassmann Numbers.}
Working with the previous ideas, we can see that for case $n=1$ in Grassmann algebra we have:
$$\mathbb{G}=alg_k\{1,x_k\}$$
that implies the polynomial expansion:
\begin{equation}
\label{n1}
p(x_1)=a+bx_1;\ \ \ a,b\in K.
\end{equation}
Also for $n=2$:
$$\mathbb{G}=alg_k\{1,x_1,x_2,x_1x_2\}$$
that also implies:
\begin{equation}
\label{n2}
p(x_1,x_2)=a+bx_1+cx_2+dx_1x_2+ex_2x_1;\ \ \ a,b,c,d,e\in K.
\end{equation}
Observe, since $\{x_1,x_2\}=0$, polynomial $p(x_1,x_2)$ can be written as:
$$p(x_1,x_2)=a+bx_1+cx_2+(d-e)x_1x_2;$$
this makes the coefficient of $x_1x_2$ behave like a non-commuting number, but in fact it is made of {\it two commuting} numbers that belongs to the field $K$.
The main tangle here was due considering the polynomial $$p(\theta)=a+b\theta$$ as part of mathematical structure, but as described above, this type of polynomial does not have a foundation on Mathematics.\\
One interesting observation is that since the Grassmann algebra $\mathbb{G}=\mathbb{G}_0\oplus \mathbb{G}_1$, the polynomial $p(x_1,x_2)\in\mathbb{G}$ can also be split as $p(x_1,x_2)=p_1(x_1,x_2)+p_0(x_1,x_2)$ where $p_1(x_1,x_2)\in\mathbb{G}_1$ takes the form: $$p_1(x_1,x_2)=bx_1+cx_2,$$ also $p_0(x_1,x_2)\in\mathbb{G}_0$ takes the form: \begin{equation}
\label{pol}
p_0(x_1,x_2)=a+(d-e)x_1x_2.
\end{equation}
It is this last equation (\ref{pol}) resembles the very initial equation (\ref{nn}) used in supersymmetry works, but now making the identifications:
\begin{eqnarray*}
a&\rightarrow&a\\
b&\rightarrow&(d-e)\\
\theta&\rightarrow&x_1x_2.
\end{eqnarray*}
where $\theta^2=x_1x_2x_1x_2=-x_1x_1x_2x_2=-0.0=0$, that agrees with usual interpretations of Berezin's and supersymmetry.
\section{\sc Conclusion.}
We saw that mathematical consistency requires the use of polynomial expansions as we have seen in eq.(\ref{n1}) or eq.(\ref{n2}) instead of that used in eq.(\ref{nn}). But as we knew from previous section, the polynomial that gives the Berezin's polynomial expansion does not have any supporting mathematical structure but is in fact {\it only part of} polynomial (\ref{n2}). One question arises: what is the meaning of the missing terms? This is not an easy question to answer since we are dealing with one supposed well know fact form supersymmetry and it will causes some discomfort to realize this misinterpretation happened, but Mathematical construction is very clear and for its formal developments we can trust on them.\\
Perhaps the following lemma will give some advice:
\begin{lemma}
	\label{pn} 
Let $p(x_1,x_2)=a+bx_1+cx_2+(d-e)x_1x_2$ be a polynomial for the Grassmann algebra $\mathbb{G}=\langle 1,x_1,x_2\rangle$, then $(p-a)^k=0$ ($\forall k>1$).
\end{lemma}
\begin{description}
\item[Proof] Is suffices to prove the $k=2$ case. We have: $$(p(x_1,x_2)-a)^2=(bx_1+cx_2+(d-e)x_1x_2)^2$$ that is of the form $$(a+b+c)^2=a^2+b^2+c^2+ab+ba+ac+ca+bc+cb,$$ observe that for this case, each squared term vanishes. This is true also for the crossed terms with $(d-e)x_1x_2$ since them will generate factors of $x_1^2$ or $x_2^2$, the only remaining terms will be $bcx_1x_2+bcx_2x_1$ that clearly vanish. So proof is completed. 
\cqd
\end{description}
The lemma have a form $Q^2=0$, that resembles the definition of nilpotent operators, used both, in Mathematics (as it is done in Operator Theory) and in Physics (e.g. Quantum Field Theory).\\  
This is, to us, the main counsel or advice, that a better understanding of non-commutative algebras (with the proper attention on PI-Theory new results) for this kind of manipulations will improve advantageous and effective results in Supersymmetry and in general in Theoretical Physics.\\
This partially clarifies the initial questions of this work about the $$b\theta$$ term of the foremost Berezin's polynomial $$p=a+b\theta,$$ wide discussed on this work, now observing coefficient $b$ not being a C-number, but constructed from a difference of elements of some field $K$ over the Grassman algebra is built.\\ But this also opens new questions to be worked out.\\
\\
{\bf Acknowledge:} Both authors are very grateful to professor German I. Gomero (at Universidade Estadual de Santa Cruz) for discussions on ideas concerning this work and also to Universidade Estadual de Santa Cruz for facilities when this work was prepared.

\end{document}